\documentclass[preprint]{aastex62}

\received{September 11, 2018}
\accepted{October 18, 2018}
\submitjournal{ApJ}

\shorttitle{Tether-cutting reconnection}
\shortauthors{Chen et al.}

\begin{document}
\title{WITNESSING TETHER-CUTTING RECONNECTION AT THE ONSET OF A PARTIAL ERUPTION}

\correspondingauthor{Hechao Chen}
\email{chc@ynao.ac.cn}
\author[0000-0001-7866-4358]{Hechao Chen}
\affil{Yunnan Observatories,Chinese Academy of Sciences, 396 Yangfangwang, Guandu District, Kunming, 650216, China}
\affil{Center for Astronomical Mega-Science, Chinese Academy of Sciences, 20A Datun Road, Chaoyang District, Beijing, 100012, China}
\affil{University of Chinese Academy of Sciences, 19A Yuquan Road, Shijingshan District, Beijing 100049, China}

\author{Yadan Duan}
\affil{Yunnan Normal University, Physics Division, Kunming 650011, Yunnan, China}

\author{Jiayan Yang}
\affil{Yunnan Observatories,Chinese Academy of Sciences, 396 Yangfangwang, Guandu District, Kunming, 650216, China}
\affil{Center for Astronomical Mega-Science, Chinese Academy of Sciences, 20A Datun Road, Chaoyang District, Beijing, 100012, China}

\author{Bo Yang}
\affil{Yunnan Observatories,Chinese Academy of Sciences, 396 Yangfangwang, Guandu District, Kunming, 650216, China}
\affil{Center for Astronomical Mega-Science, Chinese Academy of Sciences, 20A Datun Road, Chaoyang District, Beijing, 100012, China}

\author{Jun Dai}
\affil{Yunnan Observatories,Chinese Academy of Sciences, 396 Yangfangwang, Guandu District, Kunming, 650216, China}
\affil{Center for Astronomical Mega-Science, Chinese Academy of Sciences, 20A Datun Road, Chaoyang District, Beijing, 100012, China}
\affil{University of Chinese Academy of Sciences, 19A Yuquan Road, Shijingshan District, Beijing 100049, China}



\begin{abstract}
In this paper, we study the onset process of a solar eruption on 21 February 2015, focusing on its unambiguous precursor phase. With multi-wavelength imaging observations from the Atmospheric Imaging Assembly (AIA), definitive tether-cutting (TC) reconnection signatures, i.e., flux convergence and cancellation, bidirectional jets, as well as topology change of hot loops, were clearly observed below the pre-eruption filament. As TC reconnection progressed between the  sheared arcades that enveloped the filament, a channel-like magnetic flux rope (MFR) arose in multi-wavelength AIA passbands wrapping around the main axis of the filament. With the subsequent ascent of the newborn MFR, the filament surprisingly split into three branches. After a 7-hour slow rise phase, the high-lying branch containing by the MFR abruptly accelerated causing a two-ribbon flare; while the two low-lying branches remained stable forming a partial eruption. Complemented by kinematic analysis and decay index calculation, we conclude that TC reconnection played a key role in building up the eruptive MFR and triggering its slow rise. The onset of the torus instability may have led the high-lying branch into the standard eruption scenario in the fashion of a catastrophe.
\end{abstract}



\section{Introduction} \label{sec:intro}
Solar eruptions are large eruptions of magnetized plasma (up to $10^{13}$ kg)  and energy (up to $10^{32}$ ergs) from the solar atmosphere towards the interplanetary space \citep{2012LRSP....9....3W}. Because of their potential hazardous impacts on the near-Earth environment \citep{1991JGR....96.7831G,2012SSRv..171...23G} and significant disturbance at multiple solar atmospheric levels\citep{1998GeoRL..25.2465T,2012ApJ...752L..23S}, they have received considerable attention. From the inner solar atmosphere to the outer corona, a typical solar eruption often manifests as three different phenomena:  eruptive flares, filament/prominence eruption and coronal mass ejections (CMEs) \citep{1985JGR....90..275I,2013AdSpR..51.1967S}. The leading theories suggest that the three distinct phenomena can be depicted by a standard eruption scenario: the eruption of a solar magnetic flux rope (MFR) \citep{1995ApJ...451L..83S,2000JGR...10523153F}. In which, the MFR is proposed to move upward and stretch out its envelope fields when it becomes unstable somehow.  Subsequently, opposite-polarity envelope fields would come to meet and form a vertical current sheet in the wake of the rising MFR \citep{2000JGR...105.2375L}. In the current sheet, once ``flare reconnection" commences, a significant amount magnetic energy that is stored in the MFR system can be rapidly converted into kinetic and thermal energies to further power the ultimate solar eruption \citep{2006SSRv..123..251F,2015SSRv..194..237L}. To date, the existence of MFRs in CMEs has been well revealed by both solar coronal observations \citep{1999ApJ...516..465D,2011ApJ...732L..25C,2012NatCo...3E.747Z,2013A&A...552L..11L,2018ApJ...860...80P,2018ApJ...853L..18Y}, Faraday magnetic field measurements \citep{2007ApJ...665.1439L}, and even interplanetary observations \citep{2013SoPh..284..179V,2015ApJ...808L..15S,2016ApJ...817...14P}. However, how the eruptive MFR develops to deviate from its equilibrium state in the low corona still remains debated today.
\par
As the most important progenitor of solar eruptions, solar filaments often appear along the polarity inversion lines (PILs) hours or days before the onset of solar eruptions \citep{2010SSRv..151..333M}. Observations show that filaments with left-bearing barbs possess positive magnetic helicity, while filament with right-bearing barbs have negative magnetic helicity (hereafter barb rule) \citep{1998SoPh..182..107M}. Considering there are both normal-polarity (NP) and inverse-polarity (IP) filaments, two hypothetical models were proposed to describe the filament magnetic configuration. In the K-S model \citep{1957ZA.....43...36K}, a NP filament is thought to be supported above sheared arcades which refer to arcade sheared magnetic loops with the normal-polarity dips \citep{1994ApJ...430..898M,1999ApJ...510..485A}. In the K-R model \citep{1974A&A....31..189K}, an IP filament is believed to form within a MFR which refers to a group of coherent helical field lines winding one or more turn in the corona with the inverse-polarity dips \citep{1995ApJ...443..818L,2011LRSP....8....1C,2012ApJ...749..138X}.  However, the real magnetic fields of filaments might be more complex than we had thought, sometimes. With magnetic extrapolation technique, \citet{2010ApJ...714..343G} even found a coexisting of MFR and sheared arcades along a single dextral filament. More interestingly, they noticed that filament barbs in the MFR segment followed the barb rule, but filament barbs in the sheared arcade segment was against the barb rule. To clarify the correspondence between a filament barb and its magnetic configuration, \citet{2014ApJ...784...50C} proposed a more sound paradigm: filaments following (resp. against) the barb rule are formed in MFRs (resp. sheared arcades).
\par
In the past two decades, numerous trigger mechanisms for solar eruption have been proposed \citep{1980IAUS...91..207M,1989ApJ...343..971V,1999ApJ...510..485A,2000ApJ...545..524C}. In parallel with the controversial pre-eruption magnetic configuration, current main trigger mechanisms of solar eruption can also be simply divided into two categories. For ideal magnetohydrodynamics (MHD) models, a twisted MFR is routinely considered as the pre-eruption magnetic configuration \citep{2000JGR...10523153F}. Such kind of models believe a basic idea that the MFR will lose equilibrium as a critical stage was reached, involving kink instability \citep{1979SoPh...64..303H,2004A&A...413L..27T}, torus instability \citep{1978mit..book.....B,2006PhRvL..96y5002K,2010ApJ...718..433O} and catastrophic loss of equilibrium \citep{1991ApJ...373..294F,2003NewAR..47...53L} as their triggers. In particular, the torus instability of MFR may set in as the overlying envelope field decays fast enough, e.g., decay index greater then 1.5 \citep{2003A&A...406.1043T}. Note that the catastrophic loss of equilibrium of MFR is the equivalent description of ideal torus instability \citep{2010ApJ...718.1388D,2014ApJ...789...46K}. On the other hand, some other models assume sheared arcades as their pre-eruption configuration \citep{1999ApJ...510..485A,2001ApJ...552..833M}. Such kind of models typically introduce extra pre-flare reconnection below (potentially also above \citep{2010ApJ...725L..84L}) the sheared arcades resulting the formation the eruptive MFR prior to \citep{2013ApJ...764..125P,2015ApJ...809...34C,2015ApJS..219...17Y} or during \citep{2011ApJ...732L..25C,2014ApJ...792L..40S,2017NatCo...8.1330W} the related solar eruption. For instance, in the tether-cutting (TC) reconnection model \citep{2001ApJ...552..833M}, a filament is thought to be supported by the strongly-sheared core arcades that keep a magneto-static equilibrium due to the confinement of weak-sheared envelope fields. As the reconnection slowly takes place below the filament, the envelope fields that constrain the sheared core arcades would be ``cut off". At the same time, parts of the core arcades would be reconstructed into a newborn MFR moving upward and a group of low-lying small flaring loops shrinking downward. Due to the reduction of overlying confinement and the increased twist below the filament, the whole magnetic structure would expand outward and access the standard eruption scenario. A similar mechanism was proposed by \citet{1989ApJ...343..971V} in their flux-cancellation model. Compared with the TC reconnection model, the flux-cancellation model more emphasize a gradual evolution of reconnection near the photosphere. Although these trigger mechanisms have been discussed a lot from the perspective of numerical simulations \citep{2010ApJ...717L..26A,2010ApJ...708..314A,2016ApJ...832..106H,2018A&A...609A...2M}, only a very few works \citep{2014ApJ...797L..15C,2014ApJ...780...28C,2017ApJ...850...38V} provide convincing evidence to validating or distinguishing them from the perspective of observations.
\par
It is also worth noting that in the above mentioned trigger mechanism, the pre-eruption magnetic configuration of solar eruption was assumed to erupts as a whole. In fact, observations shows that the eruptive magnetic structure often undergoes a horizontal or vertical splitting and only part of its flux is expelled from the solar disk, causing a so-called partial eruption \citep{2000ApJ...537..503G,2001ApJ...549.1221G,2013ApJ...778..142T}. At present, the physical cause of partial eruption behavior is not fully understood. Assuming the pre-eruption configuration as a whole MFR, some researchers suggested that partial eruptions may set in as reconnection takes place in the interior of a filament-hosting eruptive MFR during its eruption \citep{2001ApJ...549.1221G,2006ApJ...637L..65G}. Especially for the MFR with bald patches, as reconnection occurs inside the MFR, the photospheric lines tying in the bald patches may prevent the lower part of MFR from eruption \citep{2018ApJ...856...48C}. Alternatively, other researchers tend to believed that partial erupting filaments may contain a double-decker configuration, corresponding to a double MFR equilibrium or a MFR equilibrium above sheared arcades \citep{2012ApJ...756...59L,2014ApJ...792..107K}. In such configurations, filament may split into two branches somehow and keep equilibrium for hours before the eruption of its high-lying branch. In addition, several researchers also proposed that the non-uniform magnetic twist in the MFR system might be important for its partial eruption behavior \citep{2006ApJ...645..732B,2015ApJ...805...48B}.
\par
Using observations from the Atmospheric Imaging Assembly (AIA) \citep{2012SoPh..275...17L} on board the $Solar$ $Dynamics$ $Observatory$ (SDO) \citep{2012SoPh..275....3P}, we investigate in detail the onset of a partial solar eruption. The advantage of this event is that an unambiguous and long-duration precursor phase exists before its eruption, which allows us directly to inspect its trigger mechanism and to get insights into the physical cause of its partial eruption behavior. Combined with imaging observations, kinematic analysis and decay index calculation, we find strong observational evidence that TC reconnection plays a key role in building up the eruptive MFR and triggering its slow rise. Subsequently, the torus instability appeared to take the high-lying filament branch into the standard scenario for a catastrophic eruption. The paper is structured as follows. The instruments are described in Section 2. The observations and results are described in detail in Section 3, and summary and discussion are presented in Section 4.
\section{INSTRUMENTS} \label{sec:INS}
The data we used are mainly obtained from the AIA and the Helioseismic and Magnetic Imager (HMI) \citep{2012SoPh..275..207S} on board $SDO$. The AIA uninterruptedly observes the solar atmosphere from the photosphere up to the corona through 10 EUV and UV passbands, with a temporal cadence of 12 s and a spatial resolution of  1.$^{''}$2. The HMI measures photospheric magnetic fields with 6173 \AA , and provides the full-disk line of sight (LOS) magnetograms with a temporal cadence of 45 s and a spatial resolution of 1.$^{''}$0. In this paper, we use 94 \AA \ ($\sim$6.4 MK) and 131 \AA \ ($\sim$10 MK) passbands to inspect the high temperature activity prior to and during the eruption, and also apply 171 \AA \ ($\sim$0.6 MK) and 193 \AA \ ($\sim$1.6 MK) passbands to observe its low temperature response.
Combined with the 304 \AA \ passband, we use the H$_{\alpha}$ center images from the $Global$ $Oscillation$ $Network$ $Group$ $(GONG)$ \citep{2011SPD....42.1745H} to observe the associated filament activity. Meanwhile, the 1600 \AA \ UV passband is used to reveal the reconnection signature at the lower solar atmosphere. Although active-region vector magnetic fields are available for this event, the horizontal field ($B_{h}$) is too weak to allow a reliable nonlinear force-free field extrapolation. Therefore, we investigate the photospheric magnetic field evolution beneath the filament and derive the decay index above the filament channel mainly using the LOS magnetograms. In addition, Geostationary Operational Environmental Satellite ($GOES$) X-ray data is also employed to illustrate the soft X-ray (SXR) 1$-$8 \AA \ flux of associated flare.
\section{OBSERVATIONS AND RESULTS} \label{sec:OBS}
\subsection{Overview}
\begin{figure}    
   \centerline{\includegraphics[width=1.2\textwidth,clip=]{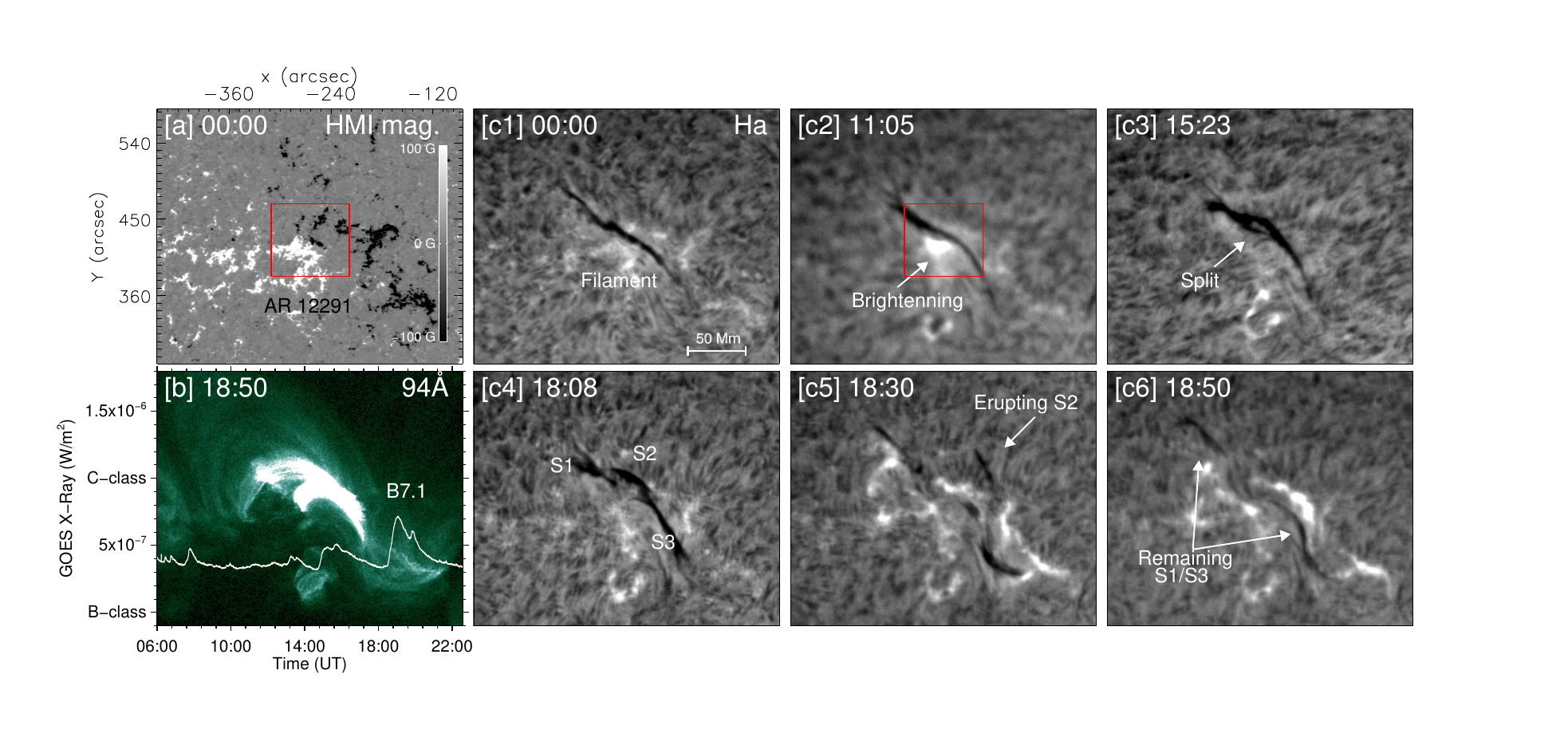}
              }
              \caption{(a) The line-of-sight magnetogram of the decayed AR 12291. (b) The B7.1-class flare caused by the partial solar eruption, in which the white curve shows the variation of the $GOES$ soft X-ray flux. (c1)-(c6) Sequential H$\alpha$ images of the activation and partial eruption of the filament, observed by the $GONG$ network. Three split filament branches are labeled as S1, S2 and S3 in panel (c4). The red boxes in panel (a) and (c2) mark the close-up field of view (FOV) of the Figure 2.}
   \label{F-1}
   \end{figure}
The solar eruption of interest occurs in the NOAA active region (AR) 12291, locating at the northern hemisphere (around N27E18). The AR 12291 is a decayed active region, and its average horizontal magnetic field is only around 60 G. Consequently, the eruption under study dose not yield a significant energy release. The eruption associates with a filament eruption and a B7.1-class flare that reaches its peak at 18:50 UT, but does not cause an obvious CME. As the progenitor of the solar eruption, the filament demonstrates some observational characteristics of a quiescent filament. For example, it has a length of $\sim$ 145 Mm, and its magnetic field strength is near 8-12 G over 15 Mm to 30 Mm based on the potential magnetic extrapolation.
\par
The main phase of the solar eruption occurs during 18:20 to 19:41 UT on 21 February 2015. Before the solar eruption gets into its eruptive main phase, a long-duration precursor phase exists from 00:00 to 18:10 UT. During the precursor phase, the filament that later erupted with the solar eruption underwent a series of typical activation phenomena \citep{2001A&A...367.1022J}. From the H$_{\alpha}$ observations, some interesting features were noticed (Figure 1(c1)-(c6)). Initially, the filament demonstrated as a straight shape and stayed still along the PIL of AR 12291. Subsequently, obvious brightennings arose right below the middle section of the filament. Afterwards, the filament slowly lifted up and displayed a series of darkenings (or widenings). Interestingly, during its activation, the filament gradually split into three distinct sections (S1, S2 and S3). By 18:08 UT, such split even became more apparent, in which S1 and S3 clearly resided in a lower height than S2. Around 20 minutes later, S2 abruptly erupt upwards somehow and yielded a flare, while S1 and S3 surprisingly remained along the PIL forming a so-called partial eruption. By inspecting these precursor activities with multi-wavelength AIA imaging observations, we are aiming to reveal the trigger mechanisms of the solar eruption, and get insight into the physical cause of its partially eruptive behavior.
\subsection{Flux Cancellation and UV brightenings}
\begin{figure}    
   \centerline{\includegraphics[width=1.\textwidth,clip=]{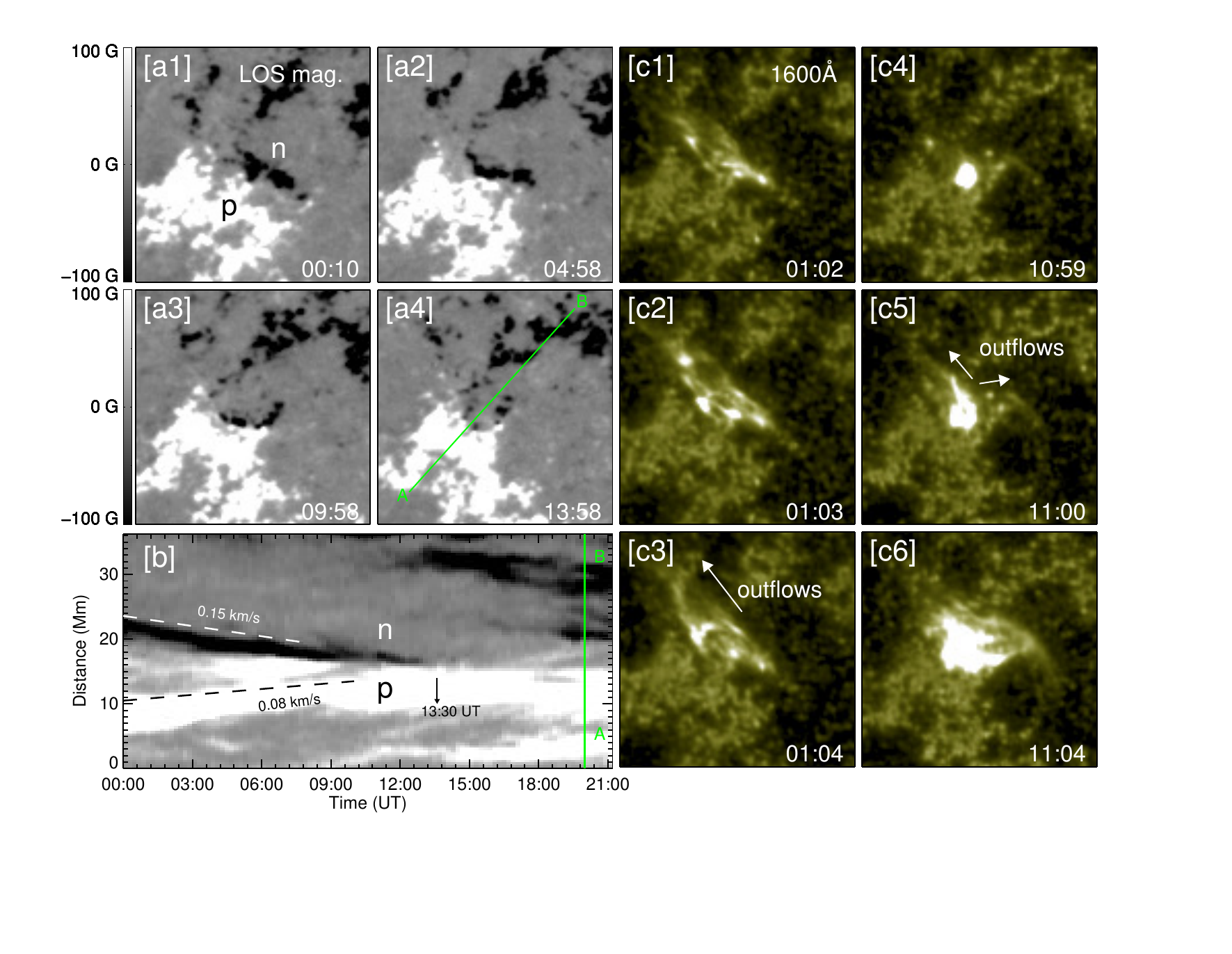}
              }
              \caption{Evidence of pre-flare reconnection beneath the middle part of the activated filament. Panels (a1)-(a4): continuous flux convergence and cancellation at the photosphere. 'p' and 'n' denote the positive and negative cancelling flux, respectively. Panel (b): the space-time stack along slice A-B in panel (a4). Panels (c1)-(c6): two episodes of UV brightennings at the chromosphere, in which outflows are marked by white arrows.}
   \label{F-1}
   \end{figure}
As mentioned above, obvious brightnenings appeared below the filament during the precursor phase of the eruption. This gives us a clue that pre-flare reconnection process might be involved in the onset of the eruption \citep{2010ApJ...725L..84L,2017ApJ...851...30X}. To verify that in detail, we carefully inspect the surface magnetic flux motion in magnetogram images and associated reconnection signatures in UV images, respectively.
\par
On the photospheric surface, we find a continuous flux convergence and cancellation occurred at the middle of the PIL. This phenomenon has been suggested as a possible manifestation of reconnection at the photosphere \citep{2018ApJ...861..135Y}. Figure 2 presents a closed-up view of such surface motions (which zoomed in the red rectangle in Figure 1 (a) and (c2)). The convergence and cancellation of opposite-polarity fluxes mainly took place along the green line ($AB$) in Figure 2(a2). We showed in Figure 5(e) the changes of unsigned negative flux in the zoomed region. The negative flux sharply decreased from 2.0 $\times$ 10$^{20}$ to 0.3 $\times$ 10$^{20}$ Mx from 00:00 to 21:40 UT. To illustrate the convergence velocity and duration of flux cancellation, a time-space stack was made along $AB$. In Figure 2(b), one can clearly see that opposite-polarity fluxes converged with a slow velocity around 0.08 $\sim$ 0.15 km s$^{-1}$, and persistently canceled for $\sim$ 13.5 hr. It is worthy of note that such obvious cancellation gradually ceased by $\sim$ 13:30 UT, but the final eruption occurred at 18:20 UT. Different from several previous observations \citep{2000A&A...356.1055J,2011A&A...526A...2G,2018ApJ...853..189P}, this feature strongly suggests that in this event, flux convergence and cancellation may be not directly responsible for the onset of the solar eruption.
\par
On the other hand, the 1600 \AA \ passband captured more convincing evidence of magnetic reconnection right above the cancelling site. In Figure 2 (c1)-(c6), we show two apparent episodes of UV plasma heating that commenced following the flux cancellation. The first episode occurred during 01:00 $\sim$ 01:08 UT, and the second episode took place during 10:58 $\sim$ 11:10 UT. Both of them started from a compact UV bright patch, and soon demonstrated as jet-like ejections associated with plasma heating. In the second episode, one can even distinguish the bidirectional outflows. Both the observations of magnetograms and UV images thus strongly imply that a pre-flare magnetic reconnection process bound to occur before the eruption.
\par
\begin{figure}    
   \centerline{\includegraphics[width=1.\textwidth,clip=]{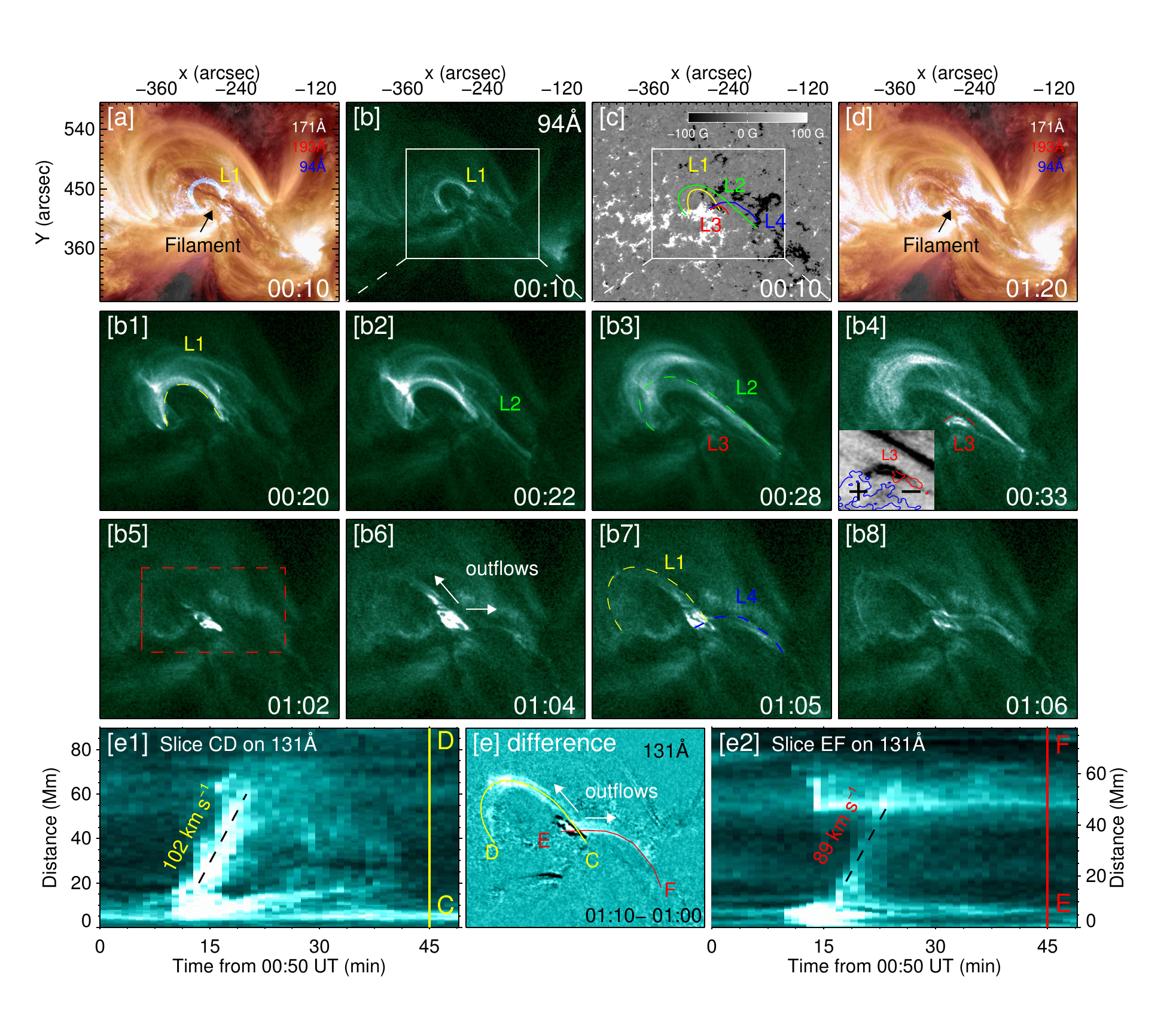}
              }
              \caption{Definitive evidence of tether-cutting reconnection. Panel (a) and (d): Composite images of AIA 171\AA, \ 193\AA,\ and 94\AA\ right before and after the first episode of reconnection. Panel (b): the close-up AIA 94\AA\ observations, in which the white window indicates the close-up FOV of panels (b1)-(b8) and (e). Panel (c): the rough configuration of magnetic loops. Panel (b1)-(b8): topology change of hot loops and bidirectional jets observed in AIA 94\AA \ images. The insert with inverse pixel values in panel (b4) demonstrates the downward-shrinking small loops, in which red/blue contour denotes negative/positive flux, respectively. The red dashed box denotes the window that we calculate the AIA flux for Figure 6(a). Panel (e): bidirectional outflows observed in AIA 131\AA \ running-difference images. Panels (e1) and (e2): the space-time stacks of 131\AA \ images along the slice CD and EF, respectively. An animation for panels (b) and (e) is available. This animation is 3 s in duration, covering 00:00:25 UT to 01:20:25 UT.}
   \label{F-1}
   \end{figure}
\subsection{Definitive signatures of Tether-cutting Reconnection}
With the EUV observations, we further investigate the pre-flare reconnection focusing on their morphology evolution, and we find that the reconnection process well agrees with the tether-cutting (TC) reconnection model of \citet{2001ApJ...552..833M}. Along with the continuous flux cancellation, there are several definitive phenomena of TC reconnection in EUV 94 \AA \ and 131 \AA \ images. These phenomena can be divided into three aspects: topology change of hot loops, the occurrence of bidirectional EUV jets, and the buildup of a channel-like MFR.
\par
\begin{figure}    
   \centerline{\includegraphics[width=1.\textwidth,clip=]{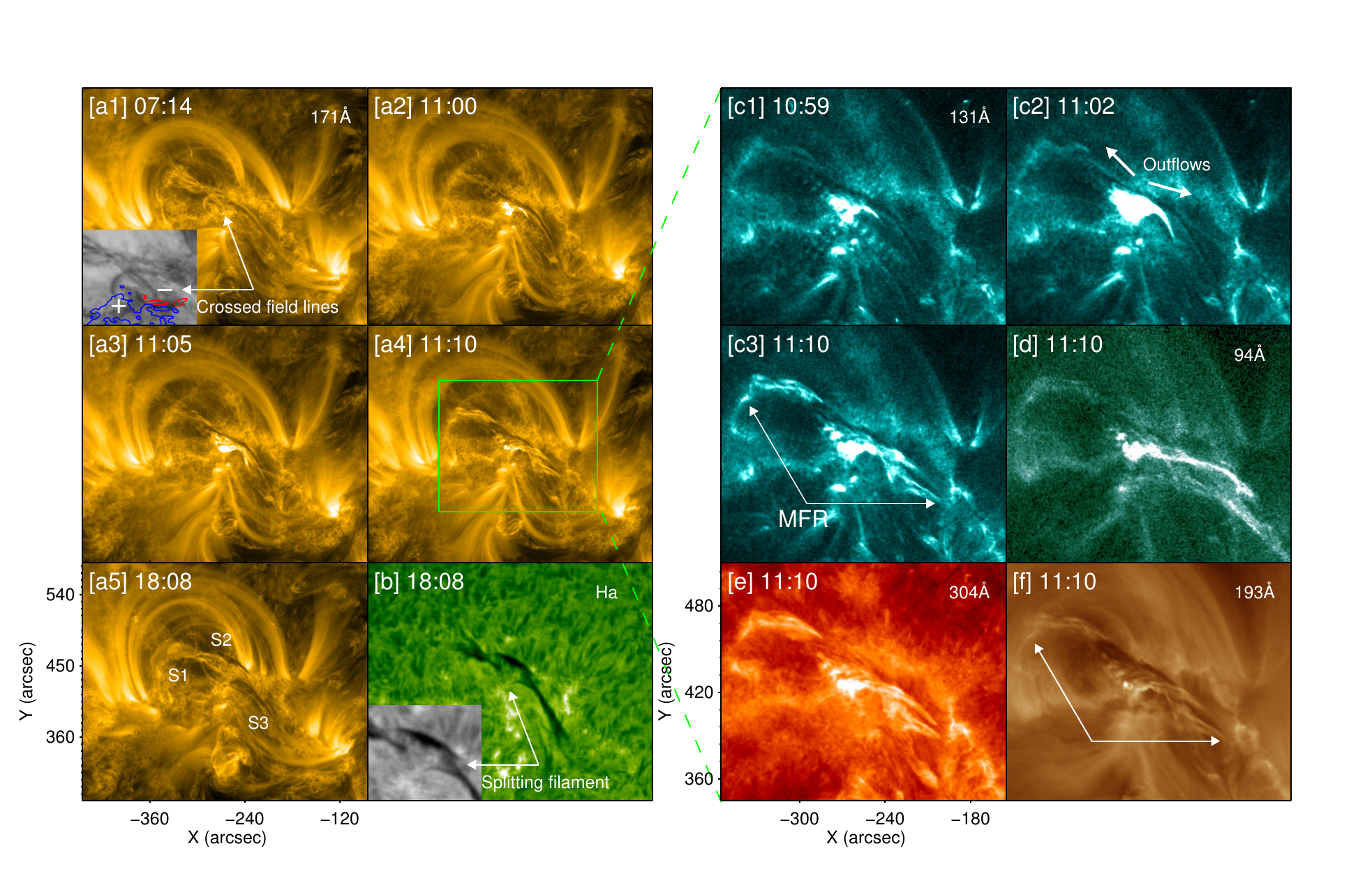}
              }
              \caption{Panels (a1)-(a5): The buildup of a channel-like magnetic flux rope in 171\AA\ images. The insert with inverse pixel values in panel (a1) illustrates the crossed field lines before the second episode of reconnection occurred, in which red/blue contour denotes negative/positive flux, respectively. Panel (b): the split filament observed by H$\alpha$ image, in which a close-up insert shows its fine structures. Panels (c1)-(c3) and (d)-(f): multi-wavelength observations of the magnetic flux rope in the green window of panel (a4). An animation of panels (a), (c), (e) and (f) is available. The buildup of the channel-like magnetic flux rope was best captured in these four AIA passbands. This animation is 5 s in duration, covering 07:10:32 UT to 10:00:32 UT.}
   \label{F-1}
   \end{figure}
\par
The topology change of hot loops happened during 00:10$\sim$00:33 UT. Initially, a group of arch-shape hot loops (L1) slowly arose astride the PIL, confining the filament. Soon L1 became brighter, and then slid its right feet down to the right. Afterwards, L1 was found to shift its right feet to a relatively remote region, leading to a newborn longer loop (L2). Interestingly, during the appearance of L2, a low-lying tiny loop (L3) also appeared below L2 after several minutes later. In the insert with inverse pixel values in Figure 3(b4), L3 was found to bridge over the cancelling site, which is very likely correspond to the downward-shrinking loops that formed during the TC reconnection.
\par
The bidirectional jet occurred during 01:00$\sim$01:10 UT, which corresponds to the first episode of UV plasma heating. Similar to its counterpart in UV observations, a compact bright patch first arose in the cancelling side, and then ejected hot plasma towards two sides, forming a so-called bidirectional jet or two-side jet \citep{2013ApJ...775..132J}.
During the jet commencement, two group of magnetic loops, L1 and L4, were well traced by bidirectional outflows. In particular, these bidirectional outflows can be discerned more easier in the running difference 131 \AA \ image (see Figure 3(e)). Along the trajectories of outflows ($CD$ and $EF$), two slices were made. From their
time-space stacks in Figure 3(e1) and (e2), one can see that the velocity of outflows along $CD$ and $EF$ were 102 km s$^{-1}$ and 89 km s$^{-1}$, respectively. The velocity of such hot outflows is roughly consistent with that of \citet{2016ApJ...818L..27C}. To better understand the topology change of hot loops and the occurrence of bidirectional jet, we outlined all the related magnetic loops, and superposed their outlines on a magnetogram (see Figure 3(c)). Thereinto, L1 and L4 refer to the sheared arcade that enveloped the filament. And the tether-cutting may be commenced between L1 and L4 due to the continuous flux convergence; L2 and L3 may refer to the newborn upward-moving long loops and the downward-shrinking small loops, respectively. Compared with Figure 3(a) and (d), it is also worthy to note that the filament indeed underwent a slow ascension with the reduction of L1.
\par
Moreover, it is found that a channel-like structure was also built up via similar TC reconnection during around 10:50 $\sim$ 11:20 UT. In Figure 4, several selected 171 \AA \ and 131 \AA \ images clearly display this process in detail. Hours before the TC reconnection commenced ($\sim$ 07:14 UT), two bundle of sheared arcades existed below the middle part of the filament. The insert image with inverse pixel values in Figure 4(a1) illustrates that these two arcades were actually rooted at opposite-polarity cancelling fluxes, enveloping the filament. As the photospheric convergence flow brought opposite-polarity magnetic elements slowly come together, TC reconnection naturally initiated between the feet of the two crossed sheared arcades. By around 10:58 UT, a compact EUV flaring patch obviously arose right at the junction of sheared arcades. Afterwards, the flaring patch started to heat and rapidly eject localized plasma towards two sides along the filed lines, leading to a bidirectional EUV jet. The occurrence of this bidirectional EUV jet, which also corresponds to the second episode of UV plasma heating we mentioned before, reveals that cool plasma around the filament was heated to a relatively higher temperature via the TC reconnection. Interestingly, the hot outflows in the bidirectional jet simultaneously well traced a newborn channel-like magnetic structure that well wrapped around the middle section of the cool filament. Note that this channel-like magnetic structure can be recognized not only in 131 \AA \ and 94 \AA \ passbands, but also in 171 \AA, \ 193 \AA, \ and 304 \AA \ passbands, suggesting this channel-like structure posses a multi-temperature feature \citep{2015A&A...580A...2Z}.
\par
To our knowledge, TC reconnection can effectively convert sheared arcades into a twisted flux rope, thus the channel-like magnetic structure should correspond to a newborn MFR. To confirm this, we carefully inspect its fine structures from a morphological perspective. Similar to the observation of  \citet{2013ApJ...770L..25L} and \citet{2014ApJ...784L..36Y}, the closer inspection in multi-wavelenth AIA observations reveals that this channel-like structure indeed consists of multi-stranded intertwined field lines (see Figure 4 (c3), (e) and (f)). In particular, such twisted fine field structures can be easily discerned in the northeast end of the channel-like structure in the close-up 131 \AA \ image of Figure 7, which supports that a MFR with apparent magnetic twist was built-up via TC reconnection during the activation of the filament. Moreover, \citet{2014ApJ...784...50C} recently proposed an indirect method to determine the configuration of filament: filaments following (resp. against) the barb rule are formed in MFRs (resp. sheared arcades) (see also \citet{2015ApJ...815...72O}). Applying this method, we find that the filament demonstrated a right-bearing barbs at its middle section, but a left-skewed overlying loop and a left-skewed dimming pattern (See Figure 7 (c) and (d)), implying the filament is an inverse-polarity configuration supported by a helical MFR. Therefore, we conclude that the channel-like magnetic structure corresponds to a newborn MFR.
\par
After its buildup, the MFR underwent a slow ascent in the following $\sim$ 7 hours, as predicted by the TC reconnection model. As a result, the original activated filament vertically split into three branches (S1, S2 and S3), in which the middle branch, S2, that wrapped around the uplifted MFR broke away from its low-lying counterparts and reached a higher height. This vertical split became most conspicuous $\sim$ 18:08 UT, and can be clearly recognized in both 171 \AA \ and H$_{\alpha}$ images (Figure 4 (a5) and (b)). The dynamic vertical split of the filament actually suggests a covert topology change of filament magnetic configuration, which well agrees with the buildup of twisted MFR via TC reconnection above two groups of sheared arcades prior to the final eruption.
\begin{figure}    
   \centerline{\includegraphics[width=1.\textwidth,clip=]{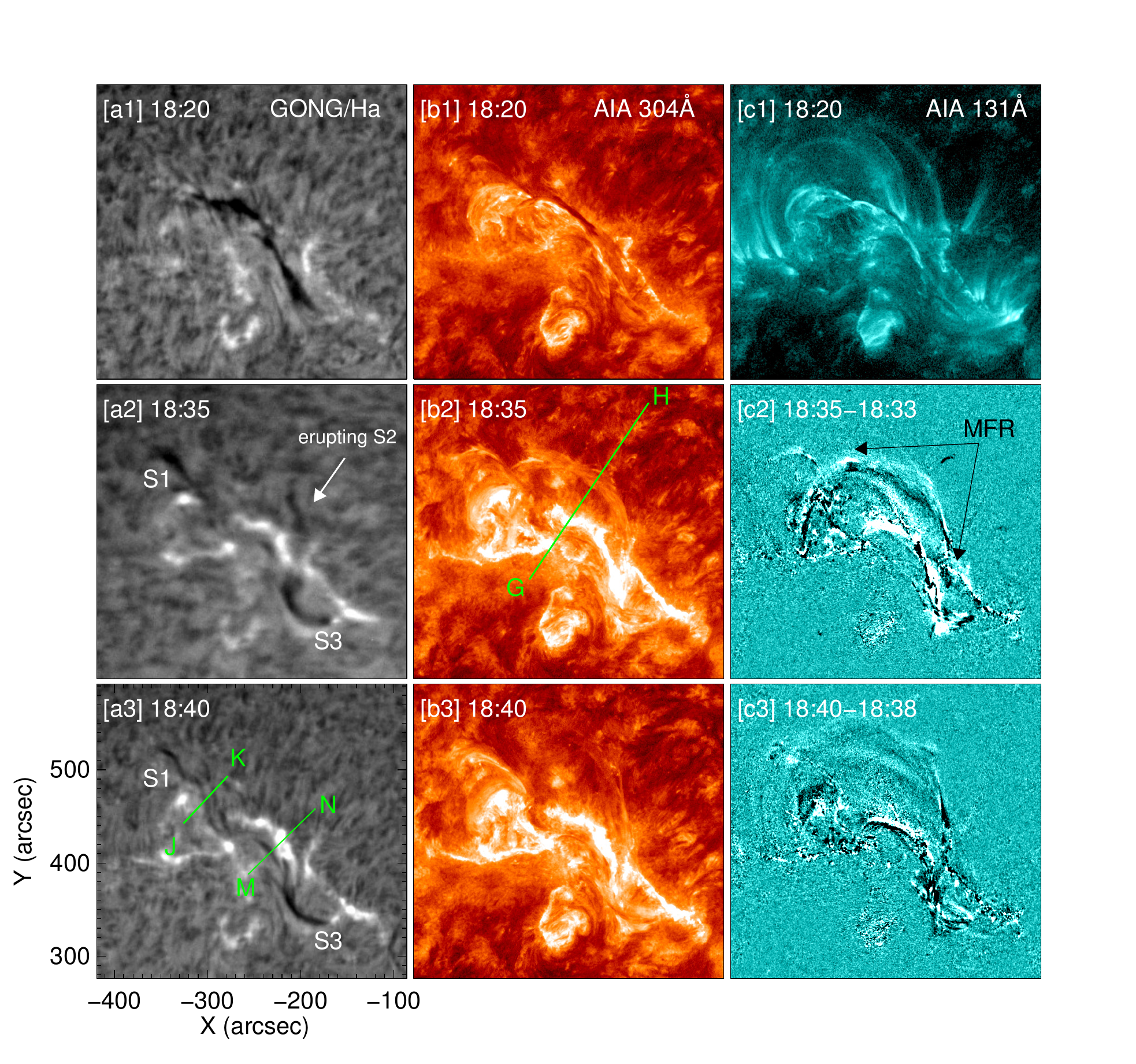}
              }
              \caption{The eruption process of the high-lying filament branch in H$\alpha$, AIA 304\AA\ and 131\AA\ images. The green slices GH, JK, and MN are aiming to trace the kinetics of the three distinctive filament branches S2, S1 and S3, respectively. An animation of this figure is available. This animation is 2 s in duration, covering 18:00:54 UT to 18:59:54 UT.}
   \label{F-1}
   \end{figure}
\subsection{The Onset of the Solar Eruption}
As S2 reached a higher height, the solar eruption soon entered its main phase. As mentioned before, this gentle eruption only yielded a B7.1-class flare, but it provides us a good opportunity to get insight into its trigger mechanism, as well as the physical cause of its partial eruption behavior. Figure 5 briefly demonstrates this eruption process with several selected H$_{\alpha}$, 304 \AA \ and 131 \AA \ images. By $\sim$ 18:20 UT, dispersive brightenings began to appear below the bifurcate filament. Meanwhile, the H$_{\alpha}$ observations shows that cool plasma in the filament became disturbed. Afterwards, the middle branch of the filament, S2, detached from its low-lying counterparts, and erupted upward. During its eruption, one can see that the erupting S2 left a flare ribbon behind itself, and the erupting filament branch in 304 \AA \ is likely to be enveloped by an relatively expended MFR in the difference 131 \AA \ images (Figure 5 (c2) and (c3)). By $\sim$ 18:40 UT, the erupting MFR became faint in EUV images, whereas the remaining filament branches, S1 and S3, became more remarkable in the H$_{\alpha}$ image. Meanwhile, two dimming regions appeared near the two ends of the eruptive MFR in 193 \AA \ image (see Figure 7 (d)).
\par
\begin{figure}    
   \centerline{\includegraphics[width=1.2\textwidth,clip=]{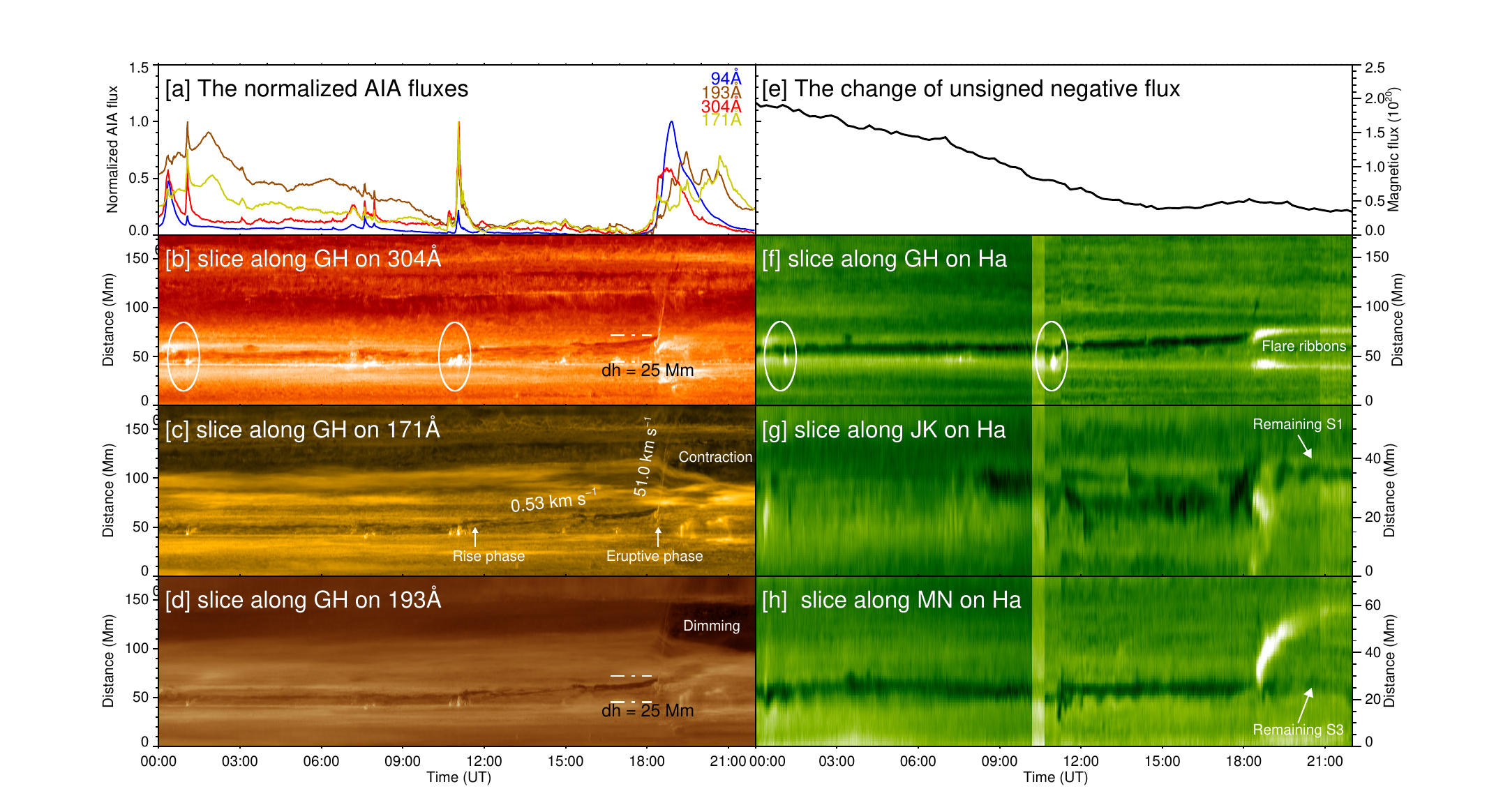}
              }
              \caption{Panel (a): the AIA fluxes change with time, calculated in the red dashed box of Figure 3(b5). Panels (b)-(d): the space-time stacks along the slice GH from 304\AA,\ 171\AA,\ and 193\AA\ images. Panel (e): the change of unsigned negative flux with time, calculated in the FOV of Figure 2. Panels (f)-(h): space-time stacks along the slice GH, JK and MN in H$\alpha$ images, respectively. The white eclipses in panel (b) and (f) denote the two episodes of reconnection mentioned above.}
   \label{F-1}
   \end{figure}
To investigate the kinematics of the partial solar eruption, we made three slices to track the dynamic behaviors of S1 (along green line, JK, in Figure 5(a3)), S2 (along green line, GH, in Figure 5(b2)) and S3 (along green line, MN, in Figure 5(a3)), respectively. We also calculated the AIA fluxes above the reconnection-cancelling site during 00:00 to 22:00 UT (within the red dashed box in Figure 3(b5)). Moreover, we calculated the change of unsigned negative flux over the cancelling site, as well. These results are presented in Figure 6. Based on the change of normalized AIA fluxes with time (Figure 6(a)), it is found that the main phase of the solar eruption started at $\sim$ 18:20 UT, and episodes of EUV flaring activities happened during the precursor phase of the solar eruption (especially from 00:00 to 12:00 UT). Thereinto, the two obvious flux peaks have been investigated in detail previously, which respectively correspond to two episodes of TC reconnection in the space-time stacks along GH on 304 \AA \ and H$_{\alpha}$ images (circled by white ellipses). The first episode of TC reconnection ($\sim$ 00:20 to 01:10 UT) triggered obvious UV brightenings, topology change of hot loops, and the bidirectional EUV jet. In addition to these reconnection phenomena, the second episode of TC reconnection ($\sim$ 10:50 to 11:20 UT) led to the buildup of a newborn MFR. Accordingly, the middle branch of the filament, S2, initiated its slow rise process with the apparent velocity of $\sim$ 0.53 km s$^{-1}$ after the second episode of TC reconnection (see Figure 6 (b)-(d) and (f)). This kind of quasi-static slow rise lasted for $\sim$ 7 hours, and is consistent with the first type of long-duration filament evolution recently reported by \citet{2018ApJ...857L..14X}.
And by $\sim$ 18:20 UT, S2 suddenly erupt upward with a higher apparent velocity of $\sim$ 51.0 km s$^{-1}$, leaving flare ribbons behind itself. This abrupt exponential increase in the upward-moving velocity of S2 is very likely to represent an occurrence of ideal MHD instability in the eruptive MFR \citep{2006PhRvL..96y5002K,2010ApJ...708..314A,2014ApJ...780...28C,2014ApJ...789L..35C}. Following this exponential acceleration process, obvious loop contraction and its related dimming can also be noticed in 171 \AA \ and 193 \AA \ passbands (see Figure 6 (c) and (d)).
\par
\begin{figure}    
   \centerline{\includegraphics[width=1.\textwidth,clip=]{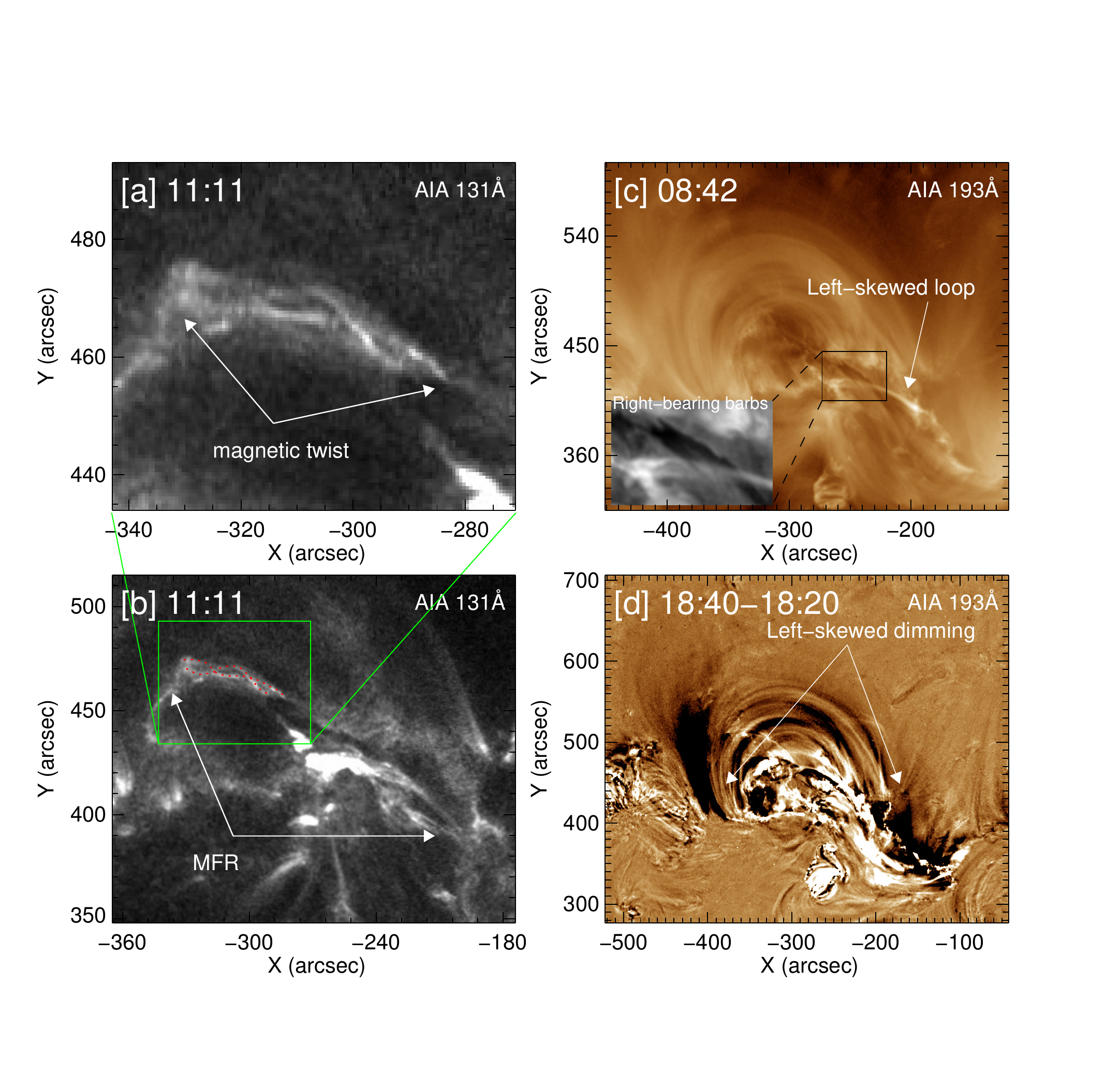}
              }
              \caption{Panels (a) and (b): the magnetic flux rope and its apparent magnetic twist in 131\AA\ image. Red dashed lines outline the intertwined fine field lines within the MFR. Panel (c) and (d): the left-skewed overlying loops, the filament right-bearing barbs and the left-skewed dimming regions are observed in 193\AA\ images. }
   \label{F-1}
   \end{figure}
As mentioned in the previous paragraph, an ideal MHD instability might play an important role in the rapid transition between the slow rise process of the erupting S2 and its exponential acceleration process. Considering the erupting S2 did not demonstrate obvious writhing/rotating motion \citep{2007ApJ...661.1260L,2012ApJ...758...42B} or a helically deformed axis \citep{2005ApJ...622L..69R,2017ApJ...850...38V} that often induced by kink instability, ideal torus instability may be the most plausible candidate. To validate this possibility, we further analyze the apparent spatial height variation of S1, S2, and S3 with time and calculate the magnetic decay index along the PIL of the filament.
From the space-time stacks in Figure 6, it is found that S2 elevated $\sim$ 25 Mm above its original height during the precursor phase of the filament. For S1 and S3, even that they underwent some visible disturbance and oscillations \citep{2014ApJ...786..151S,2017ApJ...842...27Z}, however, they eventually kept equilibrium at their original height until the eruption of S2 ended (see Figure 6 (g) and (f)).
To inspect the magnetic field situation above the eruptive filament, we first extrapolate the 3D coronal magnetic fields through the potential field source surface model (PFSS) \citep{1981A&A...100..197A,1989ApJS...69..323G}; and then determine the magnetic decay index along the PIL in detail by the formula:
$n$ $ = -d $ ln$( B_{h})/ d $ ln$(h)$ , where $B_{h}$ is the horizontal component of the background magnetic fields and $h$ is the spatial height (see results in Figure 8). Figure 8(a) displays the overall coronal magnetic field distribution over the region of our interest. In which, three branches of the filament resided along the PIL of the decayed AR 12291, well confined by bundles of overlying coronal loops. From Figure 8 (b1)-(b3), it is found that before the eruption main phase started (17:30 UT), a torus-unstable magnetic domain first appears at the middle of the PIL at $\sim$45 Mm, and then displays a two-side expansion at $\sim$60 Mm. For each filament branch, the change of its average decay index with height is also calculated at 17:30 UT in figure 8(c), respectively. Thereinto, the critical height of S1, S2 and S3 is about 64 Mm, 43 Mm and 52 Mm, respectively.
These calculations strongly reveal that a torus-instable domain indeed existed above the PIL, and the middle section of the filament might suffer from torus instability with a lower height than its counterparts. Because the filament under study is a long ($\sim$ 145 Mm) intermediate filament, its possible height should range from 15 Mm to 30 Mm \citep{2018ApJ...857L..14X}. Therefore, we believe S2 may reach the critical height ($\sim$43 Mm) and suffer the ideal torus instability after its slow levitation with the projected height of $\sim$ 25 Mm; while the low-lying S1 and S3 survived the eruption due to their lower height.
\begin{figure}    
   \centerline{\includegraphics[width=1.\textwidth,clip=]{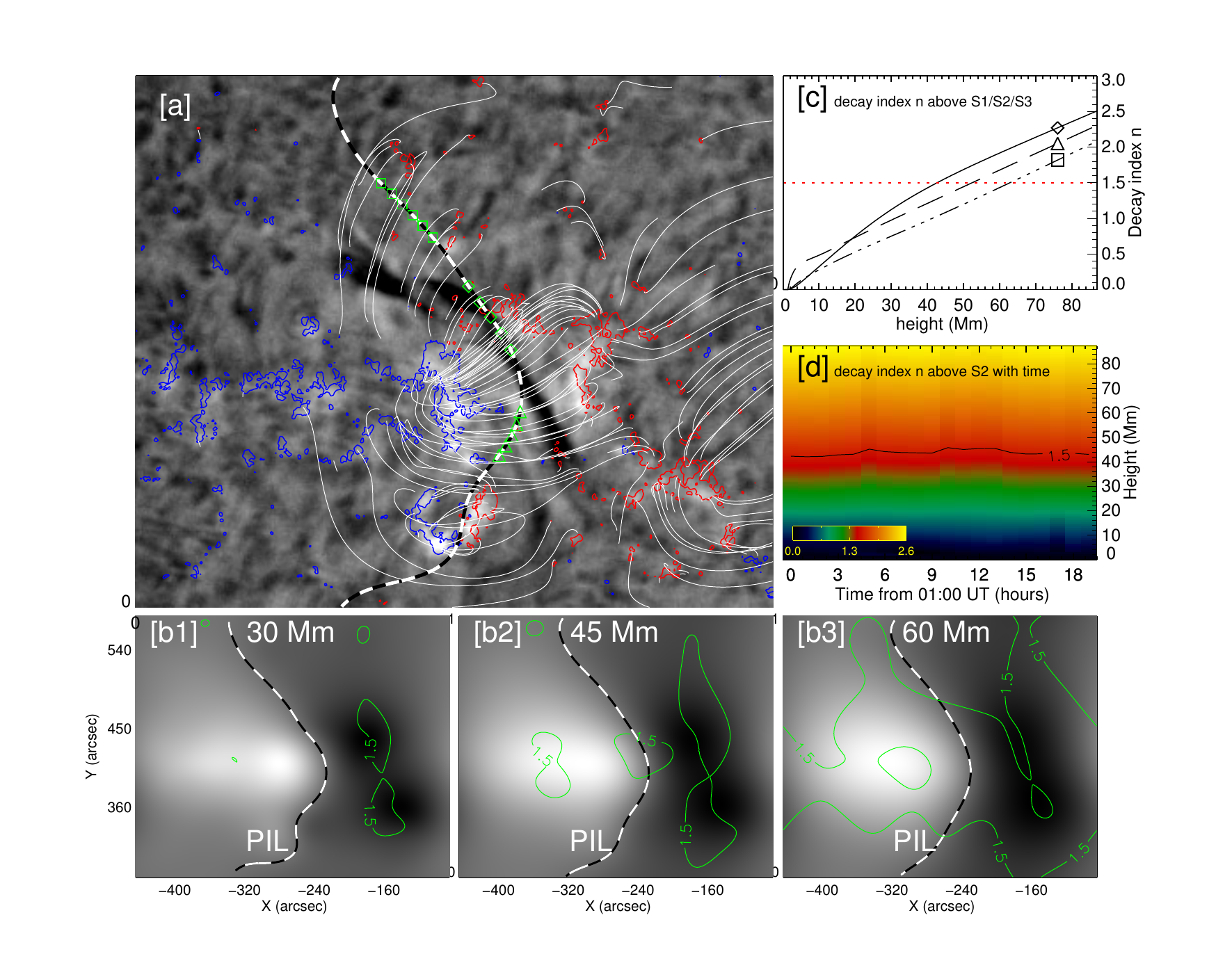}
              }
              \caption{The calculation results of magnetic decay index $n$. Panel (a): overview of the extrapolated magnetic fields in the decayed AR 12291. The black-and-white curve denotes the PIL at the height of 40 Mm; red/blue contour denotes negative/positive flux, respectively. Green squares, diamonds, and triangles mark three distinctive sections along the PIL, which respectively correspond to the rough location of S1, S2, and S3. Panels (b1)-(b3): Distributions of torus-unstable domains at various height in AR 12291, calculated at 17:30 UT, in which only contours for $n$=1.5 are plotted. PILs are plotted according to their corresponding heights. Panel (c): the change of averaged decay index with height estimated above S1, S2 and S3. Panel (d): the change of averaged decay index with time estimated above S2.}
   \label{F-1}
   \end{figure}
\subsection{The posteriori observation: the reformation of a homologous filament via similar TC reconnection}
\begin{figure}    
   \centerline{\includegraphics[width=1.\textwidth,clip=]{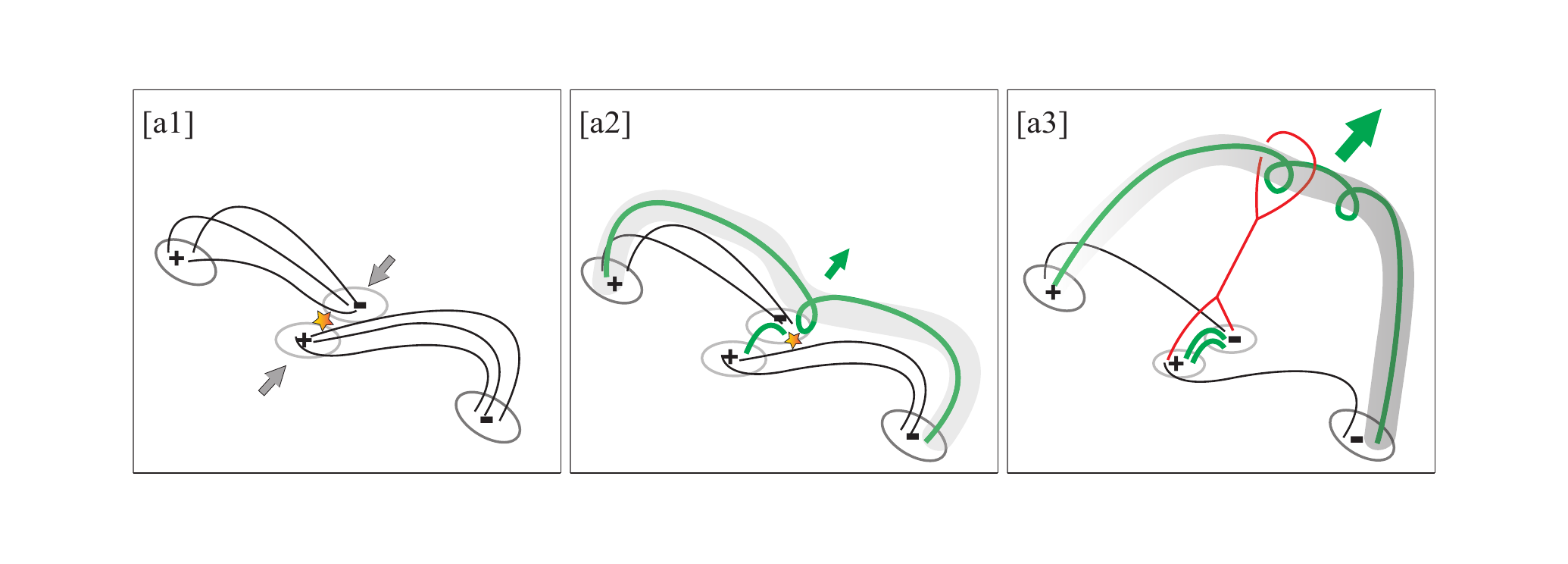}
              }
   \vspace{-2cm}
   \centerline{\includegraphics[width=1.\textwidth,clip=]{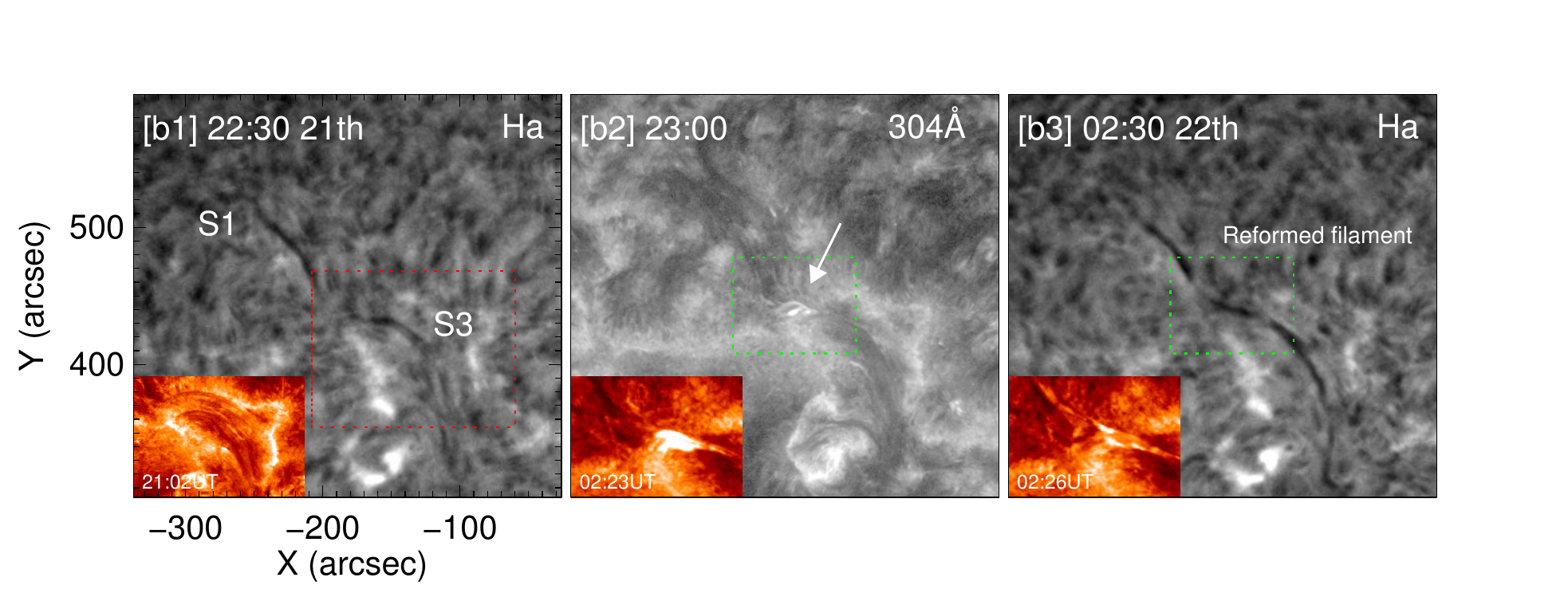}
              }
              \caption{Panels (a1)-(a3): Cartoons explaining the important aspects of our observations, which are based on the proposed tether-cutting model of \citet{2001ApJ...552..833M}. The four ellipses with +/- symbols denote the photospheric magnetic fluxes; the gray arrows denote photospheric converging flows; the black thin lines represent the sheared arcades; the yellow stars represent the reconnection; the thick green lines denote the newborn downward-shrinking loops and upward-moving helical field lines; the gray shadows represent the magnetic flux rope; the red line outlines the erupting magnetic structures. Panels (b1)-(b3): The filament reformation occurred after the eruption of S2. In panel (b1), the insert image shows that S3 consists of thread-like sheared arcades. In panel (b2) and (b3), insert images are aiming to illustrate the reconnection signatures during the buildup of the reformed filament.}
   \label{F-1}
   \end{figure}
Although the magnetic configuration of filaments remains controversial, the formation of filaments has been intensely studied for many years. To date, most observations suggested that the formation of filaments is closely related to flux cancellation and magnetic reconnection \citep{2011A&A...526A...2G,2016ApJ...827..151Y,2016ApJ...830...16Y,2017ApJ...834...42J,2017ApJ...839..128W}, but others have suggested that flux emergence are important for the formation of filaments \citep{2009ApJ...697..913O,2010ApJ...718..474L,2017ApJ...845...18Y,2018SoPh..293...93C}. For our event, the formation process of the filament was not captured during its solar disk passage, but we can infer it via the ensuing filament reformation process after the partial eruption of the high-lying eruption. Conducting a posteriori observation, here we find that the remaining low-lying filament branches, S1 and S3, reconnected with each other, causing the reformation of a homologous filament. In Figure 9 (b1)-(b3), three selected images demonstrate this process. By 22:30 UT on February 21, the remained filament branches, S1 and S3, independently existed along the PIL. These two faint and short features look like the two distinctive dark thread-like structures, as reported by \citet{2016ApJ...816...41Y}. A close-up insert in Figure 9(b1) shows that S3 actually corresponds to a bundle of sheared arcades, within which cool chromospheric plasma is responsible for its apparent darkness. In the following hours, similar TC reconnection was found to happen at their junction. During the TC reconnection, episodes of brightening signals and new magnetic connections can be observed (as illustrated in Figure 9 (b2) and (b3)). By the time of 02:30 UT on February 22, a reformed filament was fully built up. This posteriori observation not only suggests that the filament under study should also be built up via TC reconnection processes \citep{2017ApJ...840L..23X}, but also proves that S1 and S3 corresponded to two bundles of low-lying sheared arcades.
\section{SUMMARY AND DISCUSSION} \label{sec:SUM}
Imaging observations of the precursor phase of solar eruptions are to finding out the trigger mechanism of solar eruptions, and understand its complex dynamic behavior.
In this study, we investigate the onset process of a partial solar eruption in detail with multi-wavelength observations from AIA, focusing on its pre-flare reconnection phenomena, filament splitting, and trigger mechanism. This unambiguous event provides strong evidence that tether-cutting reconnection played an important role in the onset of the solar eruption. The major results of the present study are listed as follows.
\begin{enumerate}
  \item  Using multi-wavelength imaging observations, episodes of tether-cutting (TC) reconnection were observed beneath the middle part of the filament during the precursor phase of the solar eruption. This was evidenced by definitive TC reconnection signatures below (or along) the activated filament: obvious photospheric flux convergence and cancellation, episodes of UV brightenings, EUV bidirectional jets, downward-shrinking hot loops, and upward-moving helical field lines.
  \item As TC reconnection commenced between sheared arcades that enveloped the filament, a newborn magnetic flux rope (MFR) with obvious magnetic twist arose in multi AIA passbands, wrapping around the middle section of the filament. This strongly supports the idea that in this case, the eruptive MFR was built up via TC reconnection prior to the solar eruption. With the slow rise of the MFR at the velocity of 0.53 km s$^{-1}$, the middle section of the filament broke away upwards from its low-lying counterparts and reached a higher height, leading to an interesting filament vertical split.
  \item Following the slow rise phase, the high-lying branch (S2) abruptly erupted upward at the velocity of 51 km s$^{-1}$, leaving a two-ribbon flare behind; while its low-lying counterparts (S1 and S3) survived the eruption. From our kinematic analysis and decay index calculation, we suggest that TC reconnection played a key role in triggering the slow rise of the high-lying S2, and ideal torus instability may be responsible for taking the slow-rise S2 into its exponential acceleration phase.
  \item Moreover, in the posteriori observations, the remianing S1 and S3 were found to persistently reconnected with each other via similar TC reconnection. As a result, a homologous filament clearly built up by 02:30 UT on February 22. This reinforces that TC reconnection tends to be an effective and common way for the formation of filaments or eruptive MFRs in the low corona.
\end{enumerate}
\par
TC reconnection model had been proposed for a long time, however, witnessing its detailed process at the onset of solar eruption is still a challenge. As suggested by \citet{2001ApJ...552..833M}, one key reason is that the TC reconnection process tends to merge imperceptibly into the post-flare arcade reconnection. Thus, up until now, the TC reconnection model has only been supported by indirect evidence \citep{2003ApJ...599.1418S,2010ApJ...721.1579R,2013ApJ...778L..36L,2018ApJ...860..163W} or poorly observed phenomena \citep{2008ApJ...683..510K,2016ApJ...818L..27C}. In this study, we present a direct and unambiguous observation of a persistent TC reconnection process occurring in the precursor phase of a partial solar eruption. During the pre-flare reconnection process, well observed phenomena of TC reconnection were detected, i.e., continuous flux convergence and cancellation at the photosphere; obvious UV brightenings and hot outflows at the chromosphere; the appearance of EUV bidirectional jets, downward-shrinking small loops, and upward-moving newborn MFR at the low corona. Similarly, \citet{2014ApJ...797L..15C} also reported a direct observations of TC reconnection during two successive solar flares on the solar limb. Compared with their observations, our observations here not only shed more light on the close relationship between the TC reconnection and the associated filament activity on the solar disk, but also reveal some covert information on the magnetic configuration of the studied filament (see the next paragraph). As complementary, we also provide a posteriori observations, in which the reformation of a new filament via similar TC reconnection was clearly observed. These observations provide strong evidence to support the TC model of \citet{2001ApJ...552..833M}.
\par
Considering the key role of tether-cutting reconnection in the whole event, three brief illustrations are proposed in Figure 9 referring to the model of \citep{2001ApJ...552..833M}, so as to explain the important observational aspects: filament activation and split, the buildup of associated MFR and its partial eruption behavior. Thereinto, panels (a1), (a2) and (a3) respectively demonstrate the initiation of TC reconnection, the buildup of a high-lying MFR, and the eruption of the high-lying MFR. This whole process has been clearly evidenced by our observations in previous sections. Here we would like to emphasize that the magnetic configuration of the filament under study should be composed of a high-lying MFR and two groups of low-lying sheared arcades (refer to the cartoon in Figure 9(a2)). Such a special configuration can be inferred based on three observational features in our study: (i) The buildup of a newborn MFR via TC reconnection between sheared arcades. (ii) The low-lying filament branches, S1 and S3, demonstrated as two distinctive dark thread-like structures. Especially in the inset of Figure (b1), one can discern that S3 consists of several arcade-like threads. (iii) The formation of S1 and S3 as a new a new filament via similar TC reconnection in posteriori observations. Note that cool chromospheric plasma already exist within the dips of low-lying sheared arcades, as presented in Figure 9 (b1). These observations are compatible with previous observations \citep{2010ApJ...714..343G,2015ApJ...809...34C}, suggesting that the magnetic configuration of filaments cannot always be described as a simple MFR configuration or sheared configuration.
\par
Another important observational phenomenon in our event is the dynamic filament vertical split hours before the eruption of its high-lying branch. Previously, similar filament activities have been reported by several researchers \citep{2012ApJ...756...59L,2014SoPh..289..279Z,2018NewA...65....7T}. In particular, \citet{2012ApJ...756...59L} comprehensively analyzed the dynamic evolution of a filament with two separated branches. They found this ``double-decker" configuration sustained for days before the eruption of its upper branch, and explained it as two types of force-free configuration: a double MFR equilibrium or a single MFR equilibrium above a shear arcade (also see \citet{2014ApJ...792..107K}). In this study, a similar vertical split was also found hours before the eruption of the high-lying branch. From $\sim$11:00 to 18:00 UT, the upper branch slowly broke away from its low-lying counterparts, and reached a higher projected height. As evidenced in previous sections, this vertical split was caused by persistent TC reconnection beneath the middle part of the filament. Quite different from the rapid vertical split during partial eruptions that was reported by \citet{2018ApJ...856...48C}, in our case, the filament was found to split into three branches via a more quasi-static way. Moreover, the magnetic configuration of the split filament should be distinguish from the ``double-decker" configuration reported by \citet{2012ApJ...756...59L}. In our study, the high-lying branch was found to be wrapped by a newborn twisted MFR, while its low-lying counterparts actually corresponded to two distinctive sheared arcades. We conjecture the persistent TC reconnection in such a special configuration is the essential physical cause for the occurrence of its partial eruption.
\par
The flux-cancellation model recently has drawn a lot of attentions, especially in the study field of small-scale solar eruptions \citep{2017ApJ...851...67S,2018ApJ...853..189P,2018ApJ...864...68S}. In fact, it is the same as the TC reconnection model in nature, but only emphasizes a more gradual photospheric reconnection process \citep{2011LRSP....8....1C}. To our knowledge, TC reconnection can effectively convert sheared arcades into helical MFR, but cannot cause the eruption of a MFR alone. From the perspective of three-dimensional MHD simulation, \citet{2010ApJ...708..314A} analyzed at length the physical mechanisms that form a coronal MFR and later cause its eruption. They suggested that: flux cancellation and tether-cutting reconnection are key pre-eruption mechanisms for the buildup and the slow rise of a MFR, but they can not trigger solar eruption alone. Instead, it is the torus instability that causes the eruption as the slow-rise MFR reaches a critical height above the PIL.  In our event, continuous photospheric flux convergence and cancellation beneath the filament occurred mainly during 00:00 to 13:30 UT. However, the partial eruption happened at $\sim$ 18:20 UT. This implies that flux cancellation and convergence in this case was not enough to initiate the eruption \citep{2018ApJ...866....8Y}. Meanwhile, we also notice that the background fields above S2 did not demonstrate obvious change (or decay) following such flux convergence (see the change of averaged decay index with time estimated above S2, Figure 8(d)). During the time period of flux cancellation, persistent TC reconnection was initiated above cancelling site. As a result, sheared arcades gradually transformed into a twisted MFR that wrapped the middle section of the original filament. With the reduction of confinement and the increase of hoop force, the MFR slowly rose up in the following several hours, causing the filament to split vertically. The kinematic analysis of the high-lying filament branch (S2) shows that the slow rise of S2 initiated soon after the formation of MFR via TC reconnection; while the exponential acceleration of S2 abruptly started when S2 elevated an extra $\sim$ 25 Mm than its original height (probably 15 $\sim$ 30 Mm). Combining this result with the decay index calculation, we found a torus-unstable magnetic domain did indeed exist at $\sim$43 Mm above the middle part of the PIL. The evidence is in favor of the simulation results of \citet{2010ApJ...708..314A}, suggesting that the TC reconnection played a key role in triggering the slow rise of S2, whereas the torus instability probably took the slow-rise S2 into its standard eruption scenario in the fashion of a catastrophe.
\par

\acknowledgments
We thank the anonymous referee for providing detailed suggestions that helped improve
the paper. We also thank Jun Zhang and Xiaoli Yan for constructive comments, Junchao Hong, Yi Bi and Haidong Li for helpful discussion. SDO is a mission of NASAs Living With a Star Program.
This work is supported by the Natural Science Foundation of China under grants 11703084, 11633008,
11333007, 11503081, and 11703084; by the CAS programs "Light of West China" and "QYZDJ-SSW-SLH012"; and by the grant associated with the Project of the Group for Innovation of Yunnan Province.

%

\vspace{5mm}

\bibliographystyle{aasjournal}


\listofchanges

\end{document}